# Three-dimensional structure from single two-dimensional diffraction intensity measurement

Tatiana Latychevskaia

Physics Institute, University of Zurich, Winterthurerstrasse 190, 8057 Zurich, Switzerland
Current address: Paul Scherrer Institute, Forschungsstrasse 111, 5232 Villigen, Switzerland
*Corresponding author: tatiana@physik.uzh.ch

## Abstract

Conventional three-dimensional (3D) imaging methods require multiple measurements of the sample in different orientation or scanning. When the sample is probed with coherent waves, a single two-dimensional (2D) intensity measurement is sufficient as it contains all the information of the 3D sample distribution. We show a method that allows reconstruction of 3D sample distribution from a single 2D intensity measurement, at the z-resolution exceeding the classical limit. The method can be practical for radiation-sensitive materials, or where the experimental setup allows only one intensity measurement.

## Main text

## Introduction

The current trends in development of microscopy techniques aim towards imaging three-dimensional (3D) structure at the highest possible resolution. Conventional techniques for 3D structure determination require multiple measurements of the sample: in various orientations as in the case of crystallography [1], tomography [2-4], by combining crystallography and tomography as in the case of Bragg coherent diffraction imaging [5], by scanning through the sample as in confocal microscopy [6] and ptychography [7, 8], or by acquiring diffracted wave at different planes from the sample, as in the focal series method [9]. However, in some situations scanning or rotation of the samples cannot be realized because of either experimental setup limitations or due to the radiation damage sensitivity of the samples for which every measurements imposes a radiation damage that changes the structure at atomic scale. No tomography or even two exposures technique can be applied for high-resolution 3D imaging of a single molecule due to the radiation damage problem [10]. The worldwide construction of X-ray free electron lasers was initiated from the idea of diffraction imaging of individual biological macromolecules at the highest possible, atomic resolution [11]. In these coherent diffraction imaging (CDI) experiments, a single macromolecule is illuminated with an X-ray beam and its diffraction pattern is acquired in the far field [12]. The real space image of the macromolecule is reconstructed from the diffraction pattern by applying an iterative phase retrieval (IPR) algorithm [13, 14]; as a result, a 2D projection of the 3D macromolecule is retrieved [15]. Another example is the van der Waals structures or stacked 2D materials [16, 17], which are currently of high interest for their electronic properties that are defined by the stacking order. The atomic arrangement in these structures can be visualized by transmission electron diffraction microscopy and ptychography with the current resolution record of 0.4 Å [18]. Here also, the atomic distribution retrieved by these techniques is only a 2D projection of the 3D distribution without access to the information in the third dimension, such as the interlayer distances, which is a crucial parameters for characterization of the sample properties.

When sample is probed with coherent or partially coherent waves, the resulting 2D intensity distribution is formed by the constructive and destructive interference of the waves originating from the scatterers constituting the 3D sample. A single diffraction patterns contains all the information about the 3D positions of the scatterers in the sample. The non-trivial task is to extract this information. This idea was previously explored by Raines *et al*. who demonstrated anklylography - a method for extracting 3D information from a 2D diffraction pattern by using the curvature of the Ewald sphere [19]. In ankylography however the z resolution is large since it is defined by the curvature of the Ewald sphere that is typically very small. Another technique that utilizes interference for 3D information storage is holography [20, 21], which requires a well-defined reference wave. In holography, a reference wave provides solution to the missing phases which simplifies the reconstruction process. The reconstruction obtained from a hologram by simple wavefront propagation is always contaminated by the artifact out-of-focus and twin images [22]. These artifacts can be eliminated by applying IPR methods [23], however at the price of loosing 3D information, since the resulting reconstruction is a 2D projection of 3D sample distribution. In special case of very sparse samples without multiple scattering, compressive holography [24] or 3D deconvolution [25] can be employed to retrieve 3D distribution of the sample. In general, there is currently no structure reconstruction technique which allows 3D sample restoration from a 2D image and simultaneously accounts for multiple scattering effects in the sample.

Here, we propose a multislice iterative phase retrieval (MIPR) method which enables to restore the 3D sample distribution from a single-shot 2D intensity measurement, takes into account multiple scattering effects in the sample, and provides z-resolution that exceeds the classical resolution limit.

## Method

The method proposed here can be realized in the imaging schemes where the sample has a finite-size distribution so that support constraints can be applied. Such requirement is already fulfilled in most coherent diffraction imaging techniques. Holography, for example, by definition requires that the extent of the reference wave is larger than that of the object wave [21]. In CDI, diffraction pattern must be sufficiently oversampled which in real space leads to the sample being zero-padded [12, 26]. IPR methods in CDI employ masking support in the sample plane with additional constraints, as for example, transmission function being real and positive when imaged with X-rays [14, 27]. Spence *et al*. demonstrated a use of multislice approach [28] for reconstructing sample distribution from its diffraction pattern, for a specific case of sample consisting of three planes with a known sample distribution in one of the planes [29]. Multislice approach has also been utilized in ptychography as 3PIE algorithm [8], which, in opposite to the method described here, requires a sequence of diffraction patterns and applies no constraints to the transmission function distributions.

In the method proposed here, a multislice iterative phase retrieval is implemented as follows. The 3D sample is represented as consisting of *P* 2D planes: (1), (2), (3), ... (P). Each plane is characterised by its transmission function $t_p$, $p = 1...P$. When the wavefront propagates through the sample it propagates through all the *P* planes and is modified after passing each plane, as illustrated in Fig. 1. The incident wavefront in each plane is $b_p$ ("before") and the transmitted wavefront in each plane is $a_p$ ("after"). At each plane, it holds: $b_p = a_p / t_p$. $a_p$. Only $t_p$ directly describes the sample properties and can be subjected to constraints, such as a finite object size or positive absorption. Initially, two distributions are known: the distribution of the incident wavefront $b_1$ and the distribution of the measured intensity $I$. During the reconstruction, the wavefront propagation is calculated from the detector plane backward through all the sample planes, from *p* = P to *p* = 1. Then, the wavefront propagation is calculated forward, from *p* = 1 to *p* = P. This one round of wavefront propagation from detector to sample and back to detector constitutes one iteration. During the forward propagation, the distributions of the incident wavefronts at each plane $b_p$ are updated. During the backward propagation, the distributions of the transmission functions at each plane $t_p$ are updated. The reconstruction algorithm includes the following steps:

(i) The initial sample reconstruction is obtained through wavefront propagation from the detector plane back through planes (P), (P-1), ... (2), (1). In each plane, the reconstructed 2D distributions are filtered by applying constraints and thus a set of initial transmission functions is obtained $t_p^{(1)}$.

(ii) Using $t_p^{(1)}$, the wavefront forward propagation through all the planes is calculated and the complex-valued distributions of the incident wavefronts at all planes are obtained - $\tilde{b}_p^{(1)}$. The wavefront in the detector plane is then calculated - $\tilde{d}^{(1)}$.

(iii) $\tilde{d}^{(1)}$ is updated by replacing its amplitude with the measured amplitude $\sqrt{I}$, giving $d^{(1)}$.

(iv) $d^{(1)}$ is then propagated backward to the plane $P$, giving $\tilde{a}_P^{(1)}$.

(v) In plane $P$, the updated transmission function is calculated as $\tilde{t}_P^{(1)} = \tilde{a}_P^{(1)} / \tilde{b}_P^{(1)}$, where $\tilde{b}_P^{(1)}$ is obtained in (ii).

(vi) Constrains are applied to the obtained transmission function $\tilde{t}_P^{(1)}$, and the updated transmission function is $t_P^{(1)}$.

(vii) The incident wavefront in plane $P$ is calculated as $b_P^{(1)} = \tilde{a}_P^{(1)} / t_P^{(1)}$, where $\tilde{a}_P^{(1)}$ is obtained in (iv).

(viii) The obtained $b_P^{(1)}$ is propagated back to the next plane (P - 1) and the same chain of calculations is repeated starting from (v).

(ix) At plane (1), the distribution of the incident wavefront is known. For simplicity, we assume that the incident wave is a plane wave, and its distribution is $b_1 = 1$. When $b_2^{(1)}$ is propagated back to the plane (1), the obtained distribution is $\tilde{a}_1^{(1)}$ which should be equal to $\tilde{t}_1^{(1)}$, thus we set $\tilde{t}_1^{(1)} = \tilde{a}_1^{(1)}$. Next, the constraints are applied to $\tilde{t}_1^{(1)}$, giving the updated transmission function $t_1^{(1)}$. This completes the set of the updated $t_p^{(1)}$.

(x) $t_p^{(1)}$ is renamed to $t_p^{(2)}$ and the next iteration starts at (ii).

During the iterative process, the transmission functions in all planes are constrained and updated at each iteration, ultimately converging to their true distributions.

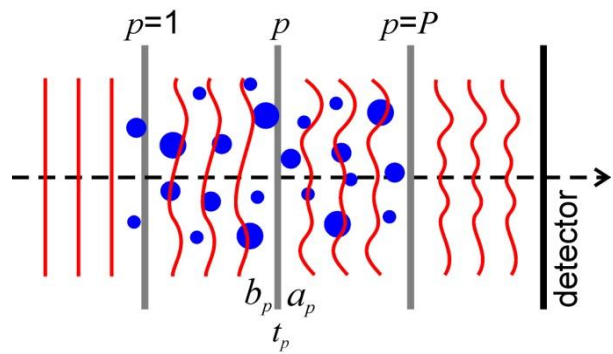

Fig. 1 Illustration of the wavefront propagation calculation and the used symbols.

The wavefront propagation can be simulated using the angular spectrum (ASM) method [30-32]. In the ASM, the wavefront distribution $u_p$ in plane p is obtained from the wavefront $u_{p-1}$ in plane (p-1) by $u_p = \text{FT}^{-1}\left[\text{FT}(u_{p-1})\exp\left(\frac{i2\pi}{\lambda}\Delta z\sqrt{1-f_x^2-f_y^2}\right)\right]$, where FT and FT$^{-1}$ are Fourier and inverse Fourier transforms, respectively, $(f_x, f_y)$ are the coordinates in the Fourier plane, $\lambda$ is the wavelength, and $\Delta z$ is the distance between the two planes. In the ASM, $\Delta z$ can be arbitrary small, for $\Delta z = 0$ the wavefront simply remains unchanged, $u_p = u_{p-1}$. Moreover, for a sample that consists of two (or more) parts positioned in the same plane ($\Delta z = 0$) and described by transmission functions $t_p^{(1)}$, $t_p^{(2)}$..., the wavefront behind that plane can be calculated by using the ASM as described above and the resulting distribution will be the same as propagation through one plane with total transmission function $t_p = t_p^{(1)} t_p^{(2)}...$

## Simulated examples

The first example is 3D reconstruction by MIPR from a single-shot in-line hologram. The 3D sample consists of four opaque objects in the form of letters positioned at four different planes, with the transmission function at each plane being $t = 0$ at the object's position and $t = 1$ elsewhere. The distances between the planes is 50 μm and the distance from the last plane to the detector is 200 μm. The hologram (Fig. 2(a)) was simulated with an incident plane wave of 532 nm wavelength by the ASM [30-32]; hologram's size was 400 μm × 400 μm. In all simulated examples demonstrated in this study (holograms and diffraction patterns), Gaussian-distributed noise was superimposed onto the intensity distribution giving the resulting signal-to-noise ratio (SNR) of 10. The reconstruction obtained by conventional wavefront backward propagation [31] is shown in Fig. 2(b). The 3D reconstruction obtained by the MIPR is shown in Fig. 2(c), the recovered 2D distributions of the transmission functions are free of out-of-focus signal and twin images artefacts. MIPR was performed for 1000 iterations, with the following constraints applied to the transmission functions: the phase was set to zero and the amplitude threshold at one, thus the positive absorption filter was applied [22]. Additionally, in planes (1), (2), and (3) a support constraint was applied in the form of a loose mask about four times larger than the object, the values outside the mask were set to one and remained unchanged inside the mask. It was observed that for a sample consisting of 2 planes no mask support was required, and for a sample consisting of 4 planes, masking support was required in 3 planes. Remarkably, reconstructions of the same quality (not shown here) were obtained when the distance between the four planes was set to $\Delta z$ = 1 μm, thus demonstrating that the z-resolution of the reconstructed 3D distribution can exceed the resolution given by the classical limit [33]: $R_{\text{Axial}} = \frac{2\lambda}{\text{NA}^2} = 3.192$ μm, where NA is the numerical aperture of the optical setup.

In the next example (Fig. 2(d)), the sample consists of four phase objects: spheres of 20 μm in diameter. The transmission function of each sphere is given by $t(x,y) = \exp[-a(x,y)]\exp[i\varphi(x,y)]$, with a maximal absorption $a(x,y)$ of 0.1 au and a maximal phase shift $\varphi(x,y)$ of 0.5 radian. The other parameters are the same as in the example above. Obtaining a quantitatively correct reconstruction of a phase object from its in-line hologram is a known problem which can be resolved only by applying IPR methods [34]. The reconstructions obtained by conventional wavefront backward propagation [31] are shown in Fig. 2(e). Here the recovered phase shift varies from -0.43 to 0.49 radian. The 3D reconstruction obtained by the MIPR is shown in Fig. 2(f), the obtained phase distributions exhibit out-of-focus and twin-image-free reconstructed spheres with quantitatively correct recovered phase shifts of 0.5 radian. The following constraints were applied to the transmission functions in the MIPR: tight mask supports for each object (sphere) were created from the initial reconstruction, the values outside the masks were set to one and remained unchanged inside the masks.

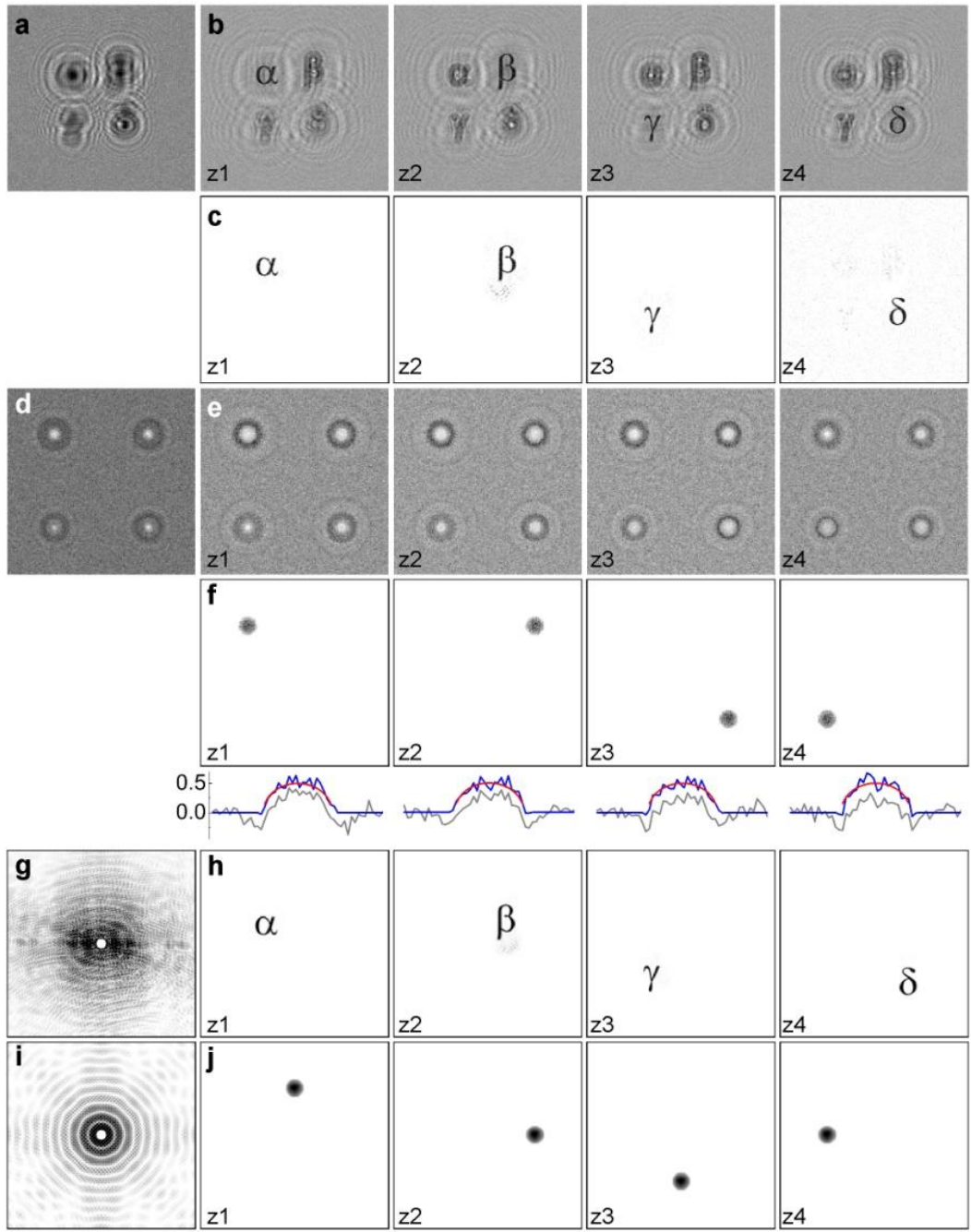

Fig. 2 Simulated examples of 3D reconstructions from a single 2D intensity measurement. 3D samples consist of individual objects located in four different planes. (a) Hologram of a sample consisting of opaque objects in the form of letters. Reconstruction obtained by (b) conventional wavefront backward propagation, and (c) the MIPR. (d) Hologram of a sample consisting of phase spheres. Phase distributions reconstructed by (e) conventional wavefront propagation and (f) the MIPR. The phase range in (e) is from -0.43 to 0.49 radian, and (f) from 0 to 0.7 radian. The plots in (f) show phase profiles of the spheres: original distributions (red), reconstructed by conventional reconstruction (grey) and by the MIPR (blue). (g) Diffraction pattern of 3D

sample made of opaque objects in form of letters (shown in logarithmic scale) and (h) its 3D reconstructions. (i) Diffraction pattern of a 3D sample consisting of phase spheres (shown in logarithmic scale) and (j) the reconstructed phase distributions. All distributions are sampled with 400 × 400 pixels, and only central regions of 200 × 200 pixels are shown, the diffraction patterns in (g) and (i) are shown in full.

Another application of the MIPR method is 3D sample reconstruction from a 2D diffraction pattern. In CDI [12, 26], a diffraction pattern of the sample is acquired in far field, and the missing phase information together with the sample distribution is retrieved by applying an IPR algorithm which runs wavefront propagation back and forth between the sample and the detector planes. Conventional IPR methods reconstruct a 2D projection of a 3D sample: $s'(x,y) = \int s(x,y,z)\mathrm{d}z$. In the MIPR method, the resulting reconstruction is not a 2D projection of the sample but it is a 3D distribution of the sample $s(x,y,z)$. Simulated examples of MIPR applied for 3D sample reconstruction from a single 2D diffraction patters are shown in Fig. 2(g) - (j). The simulation parameters were the same as for the holograms, but the wave after passing the last sample plane was propagated to the detector plane in the far field. The central region in the diffraction patterns was set to zero to mimic experimental conditions (beamstop). The initial reconstruction was obtained as follows: the complex-valued wavefront at the detector plane was formed by combining the known amplitude with a random phase distribution, and the obtained wavefront was then propagated back from the detector plane to each of the four planes. In the iterative routine, the following constraints were applied to the transmission functions: the amplitude was threshold at one, and a support constraint was applied in all planes in the form of a loose mask about four times larger than the object. For opaque objects, in addition, the phase of the transmission function was set to zero. 100 reconstructions were obtained, the one with the minimal error is shown in Fig. 2(h) exhibiting artefact-free reconstructed transmission functions at each plane. The error was calculated as $\sum_{i,j}^{N} |F_{\text{exp}}(i,j) - F_{\text{it}}(i,j)| / |F_{\text{it}}(i,j)|$, where $F_{\text{exp}}(i,j)$ and $F_{\text{it}}(i,j)$ are the experimental and measured amplitudes, respectively, and $(i,j)$ are the pixel indices; the pixels in the central region with missing intensity values were excluded from the summation.

In another example, a sample consisting of four phase spheres with a maximal absorption $a(x,y)$ of 0.1 au and a maximal phase shift $\varphi(x,y)$ of 0.5 radian positioned at non-equidistant z-distances (50 μm, 40 μm and 30 μm, respectively), was simulated. The corresponding diffraction pattern is shown in Fig. 2(i). The reconstructed phase distributions are shown in Fig. 2(j), exhibiting the four spheres with quantitatively correct recovered phase of 0.5 radian. In the iterative routine, the following constraints were applied to the transmission functions: (1) a loose mask support for each object (sphere), (2) the amplitude was threshold at one and the phase distribution was not constrained.

Simulated electron (80 keV) diffraction pattern of bilayer twisted graphene (twist angle 7°, interlayer distance 3.35 Å) is shown in Fig. 3(a). The simulations were done as explained elsewhere [35, 36], with the transmission functions of each layer given by $t^{(1,2)}(x,y) = \exp\left[i\sigma V_z^{(1,2)}(x,y)\right]$, where 1,2 is the layer number, $\sigma$ is the interaction parameter, and $V_z^{(1,2)}(x,y)$ is the projected potential of the sample, no weak phase approximation was applied. The reconstructed phase distributions of the two layers separated by 3.35 Å obtained by the MIPR are shown in Fig. 3(c) and (d). The second layer was modelled with defects, which were correctly reconstructed, Fig. 3(d). In the MIPR, the initial phase distributions were that of perfect lattices and the constraints applied to the transmission functions were: amplitude was set to one and the phase was threshold at 0.5 radian, the sample region was limited to a round patch of 4 nm in diameter.

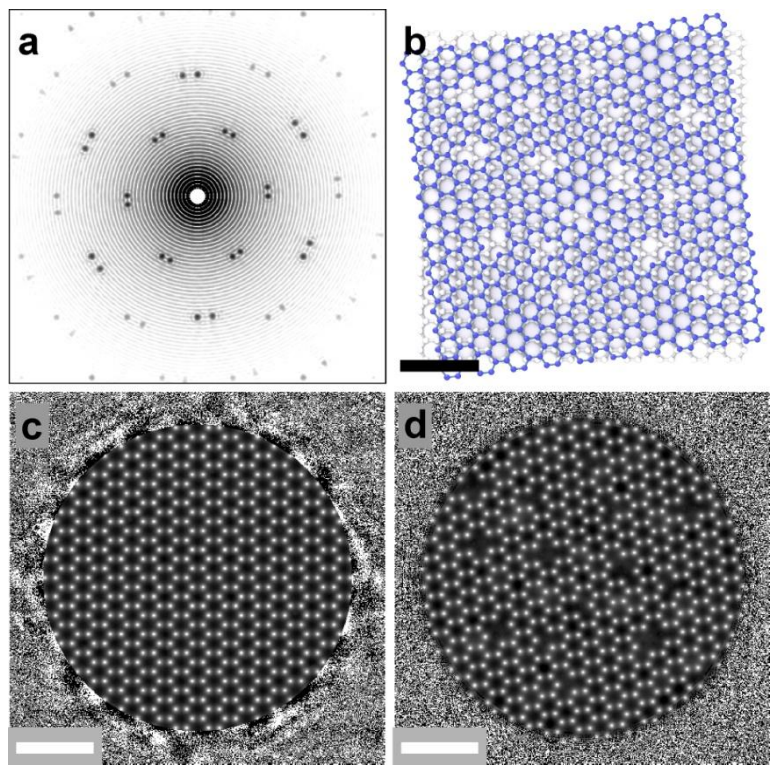

Fig. 3 Simulated 2D electron diffraction pattern of bilayer twisted graphene (BLG) shown in inverted logarithmic scale (a). (b) Modelled BLG distribution with defects in one layer. (c) and (d) The phase distributions of the reconstructed transmission

functions of the individual layers, values range from 0 to 0.24 rad. Scalebar in (b) - (d) is 1 nm.

The initial $z$-positions of the sample planes and the mask supporting constraints can be easily obtained for holography by simple back propagation of the wavefront. For diffraction patterns, they can be guessed from an approximate sample distribution, or even obtained by back propagation of the wavefront from the diffraction pattern similar to how it is done in photoelectron emission [37]. The precise $z$-positions of the sample planes can be refined from the reconstruction obtained by selecting z-positions of the sample planes which result in the minimal error. To test this hypothesis, the diffraction pattern of four opaque objects in form of letters, was reconstructed by the MIPR at incorrect distances between the planes: 50, 50 and 40 μm instead of 50, 50 and 50 μm. One hundred reconstructions were obtained for correct and incorrect z-distances and for each reconstruction the error was calculated. For the correct z-distances the minimal error was 1.134 and for the incorrect z-distances it was 1.139. For noise-free diffraction patterns with the signal in the central region available, the minimal error for the correct z-distances was 0.044 and for the incorrect z-distances it was 0.163. Thus, the precise z-positions of the sample planes can be obtained by finding the reconstruction with the minimal error. In addition, a 3D sample with zero distance between the planes 3 and 4 was simulated, the reconstructions obtained by the MIPR were identical to those in the case of finite z-distances as shown in Fig. 2(h) - all four distributions were clearly reconstructed. This demonstrates that the MIPR with wavefront propagation by the ASM can be applied for as small distances between the planes as zero, which has potential for high z-resolution imaging.

## Experimental results

For experimental proof of concept, an optical in-line hologram of a 3D sample was recorded with plane wave (532 nm) as illustrated in Fig. 4(a). The sample consisted of polystyrene microspheres of 4 μm in diameter which were prepared in a water solution and randomly distributed on both sides of a microscopic glass. The imaged area was 550 μm × 550 μm.

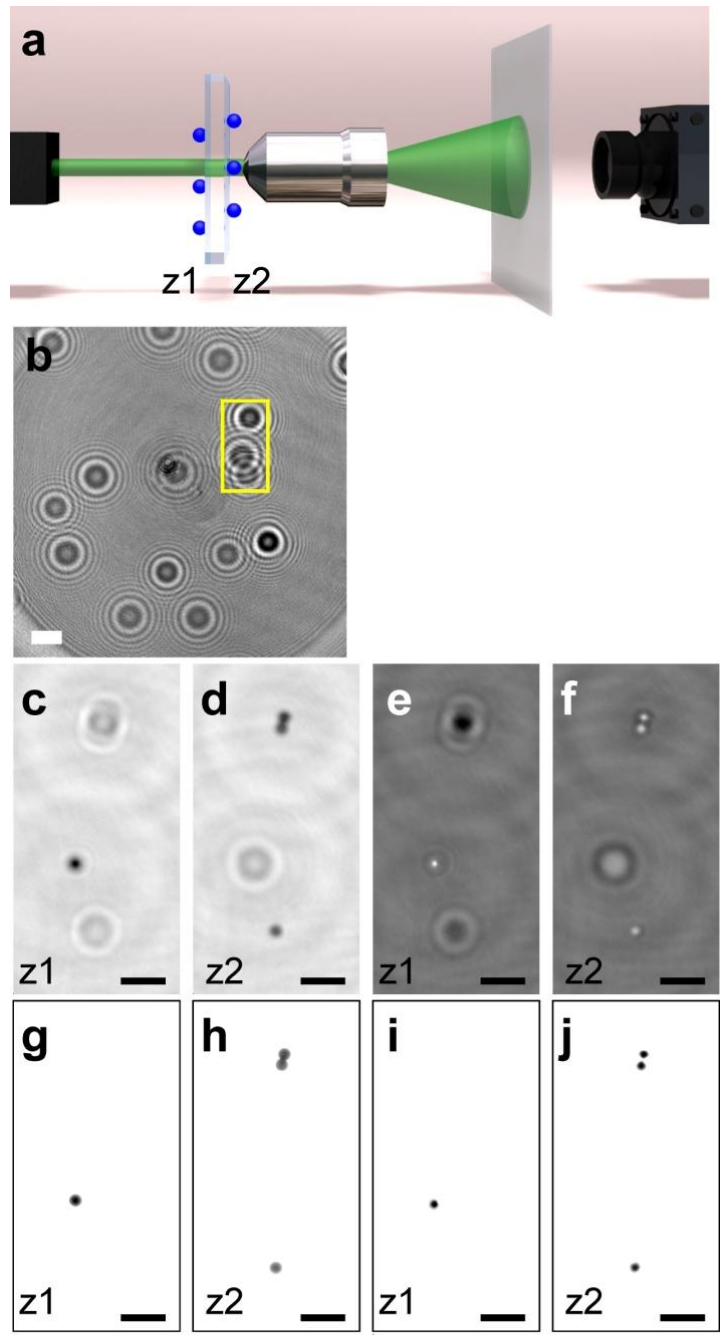

Fig. 4 Experimental results. (a) Experimental optical setup. (b) Hologram, scalebar is 50 μm. (c) - (f) Reconstructions obtained by conventional wavefront propagation, (c) - (d) amplitude and (e) - (f) phase distributions at two different z-distances. Amplitude ranges from 0.116 to 1.190 au, and phase ranges from -0.561 to 1.204 rad. (g) - (j) Reconstructions obtained by the MIPR, (g) - (h) amplitude and (i) - (j) phase distributions (inverted contrast) at the two $z$ distances. Amplitude ranges from 0.119 to 1.000 au, and phase ranges from 0.000 to 1.371 rad. (c) - (j) Reconstructions of the region marked by the yellow rectangle in (b) are shown, scalebar is 20 μm.

A conventional reconstruction was obtained by wavefront propagation from the hologram plane for a range of *z*-distances. Two sets of spheres were found in-focus at two distances counted from the hologram plane: *z*2 = 636 μm and *z*1 = 780 μm. Thus, the *z*-positions of two planes of the microscopic glass were identified. The reconstructed amplitude and phase distributions at both planes are shown in Fig. 4(c) - (f). Next, the MIPR was performed with the following constraint: tight mask supports were applied for each object in each plane, masks were created from the initial reconstruction. The reconstructed transmission functions are shown in Fig. 4(g) - (j): both sets of spheres show up each in their respective planes without out-of-focus and twin images.

## Discussion and conclusion

In conclusion, when 3D sample is probed with coherent radiation, a single-shot 2D intensity distribution measured in far field contains the 3D information about the sample distribution, which can be restored by the MIPR method. Compared with previous IPR methods limited to 2D reconstruction, the MIPR method accounts for multiple, or dynamical, scattering in the sample. The proposed iterative algorithm uses alternating reconstruction of transmission functions and incident wavefronts during back and forward propagations. The lateral resolution of the reconstructed 3D sample distribution is governed by the classical Abbe criterion, while the z-resolution can be beyond the classical criterion when employing the property of the ASM which allows for wavefront propagation for arbitrary small z distances. The proposed method opens a possibility for 3D sample reconstruction from 2D diffraction patterns in techniques which do not allow multiple measurements by scanning or rotation of the samples, either due to experimental setup limitations or radiation damage sensitivity of the samples.

## Acknowledgements

The author acknowledges support by the University of Zurich.